\documentclass[sigconf]{acmart}

\usepackage{graphicx}
   \graphicspath{{./Figures/}}

\usepackage{amsmath}    
\usepackage{algorithm}
\usepackage{algorithmic}
\usepackage{multirow}
\usepackage{svg}
\usepackage{color}

\DeclareGraphicsExtensions{.pdf,.jpeg,.png,.fig, .emf}
   \usepackage{subfigure}
    \usepackage{epstopdf}
    \usepackage{epsfig}
    \usepackage{subfigure}
    \usepackage{epstopdf}
    \usepackage{epsfig}

\copyrightyear{2024}
\acmYear{2024}
\setcopyright{acmlicensed}\acmConference[DAC '24]{61st ACM/IEEE Design Automation Conference}{June 23--27, 2024}{San Francisco, CA, USA}
\acmBooktitle{61st ACM/IEEE Design Automation Conference (DAC '24), June 23--27, 2024, San Francisco, CA, USA}
\acmDOI{10.1145/3649329.3655989}
\acmISBN{979-8-4007-0601-1/24/06}

\begin{document}
%
\title{\vspace{-10mm} HyCiM: A Hybrid Computing-in-Memory QUBO Solver 
for General Combinatorial Optimization Problems\\ with Inequality Constraints }

 \author{\small{}
         Yu Qian$^1$, 
         Zeyu Yang$^1$, 
         Kai Ni$^2$, 
         Alptekin Vardar$^3$,
         Thomas K{\"a}mpfe$^3$ 
         and Xunzhao Yin$^{1,4*}$ 
          \\$^1$
          Zhejiang University, Hangzhou, China;
          $^2$University of Notre Dame, Notre Dame, USA;
          $^3$Fraunhofer IPMS, Dresden, Germany
          \\$^4$Key Laboratory of CS\&AUS of Zhejiang Province, Hangzhou, China;
          $^*$Corresponding author email: xzyin1@zju.edu.cn
}

\renewcommand{\bibfont}{\scriptsize}

\let\oldbibliography\thebibliography
\renewcommand{\thebibliography}[1]{\oldbibliography{#1}
\setlength{\itemsep}{-0.5pt}} 


\begin{abstract}
\vspace{-1ex}
Computationally challenging combinatorial optimization problems (COPs) play a fundamental role in various applications.
To tackle COPs, many Ising machines and Quadratic Unconstrained Binary Optimization (QUBO) solvers have been proposed, which typically involve direct  transformation of COPs into Ising models or equivalent QUBO forms (D-QUBO). 
However, 
when addressing COPs with inequality constraints, this D-QUBO approach introduces numerous extra auxiliary variables, resulting in a substantially larger search space, increased hardware costs, and reduced solving efficiency.
In this work, we propose HyCiM,  a novel hybrid computing-in-memory (CiM) based QUBO solver framework, designed to overcome aforementioned challenges. 
The proposed  framework consists of 
(i) an innovative transformation method (first to our known) that converts COPs with inequality constraints into an inequality-QUBO form, thus eliminating the need of expensive auxiliary variables and associated calculations; 
(ii) 
"inequality filter",  a ferroelectric FET (FeFET)-based CiM circuit  that accelerates the inequality evaluation, and filters out infeasible input configurations; (iii) 
a FeFET-based CiM annealer that is capable of approaching global solutions of COPs via iterative QUBO computations within a 
simulated annealing process.
The evaluation results show that HyCiM drastically narrows down the search space, eliminating $2^{100} \text{ to } 2^{2536}$ infeasible input configurations compared to the conventional D-QUBO approach.
Consequently, the narrowed search space, reduced to $2^{100}$ feasible input configurations, leads to a substantial hardware area overhead reduction, ranging from 88.06\% to 99.96\%.
Additionally, 
HyCiM consistently exhibits a high solving efficiency, achieving a remarkable average success rate of 98.54\%, whereas D-QUBO implementatoin shows only 10.75\%.
\end{abstract}

\maketitle
\pagestyle{empty}

%

\vspace{-2ex}
\section{Introduction}
\label{sec:intro}
\vspace{-1ex}


Combinatorial optimization problems (COPs) find applications in various domains, including logistics, resource allocation, communication network design, 
and transportation systems, among others 
\cite{paschos2014applications, chang2022analog}.
Many of these problems are classified as non-deterministic polynomial-time hard (NP-hard), indicating their complexity within the NP realm.
A wide range of COPs, including Max-Cut, graph coloring, and knapsack problems, can be effectively reformulated into Ising models or Quadratic Unconstrained Binary Optimization (QUBO) forms, which can be addressed by Ising machines and QUBO solvers 
\cite{yin2024ferroelectric, mohseni2022ising}. 
%
%
For example, digital ASIC annealers
\cite{katsuki2022fast,takemoto20214} 
implement different annealing algorithms within digital circuits, while dynamical system Ising machines \cite{moy20221, ahmed2021probabilistic} utilize the intrinsic system dynamics and their tendency to settle at ground state to solve COPs.
These solvers are specifically designed to solve COPs much more quickly and efficiently than von Neumann hardware.
%

\begin{figure}[!t]
  \centering
  \includegraphics[width=0.9\columnwidth]{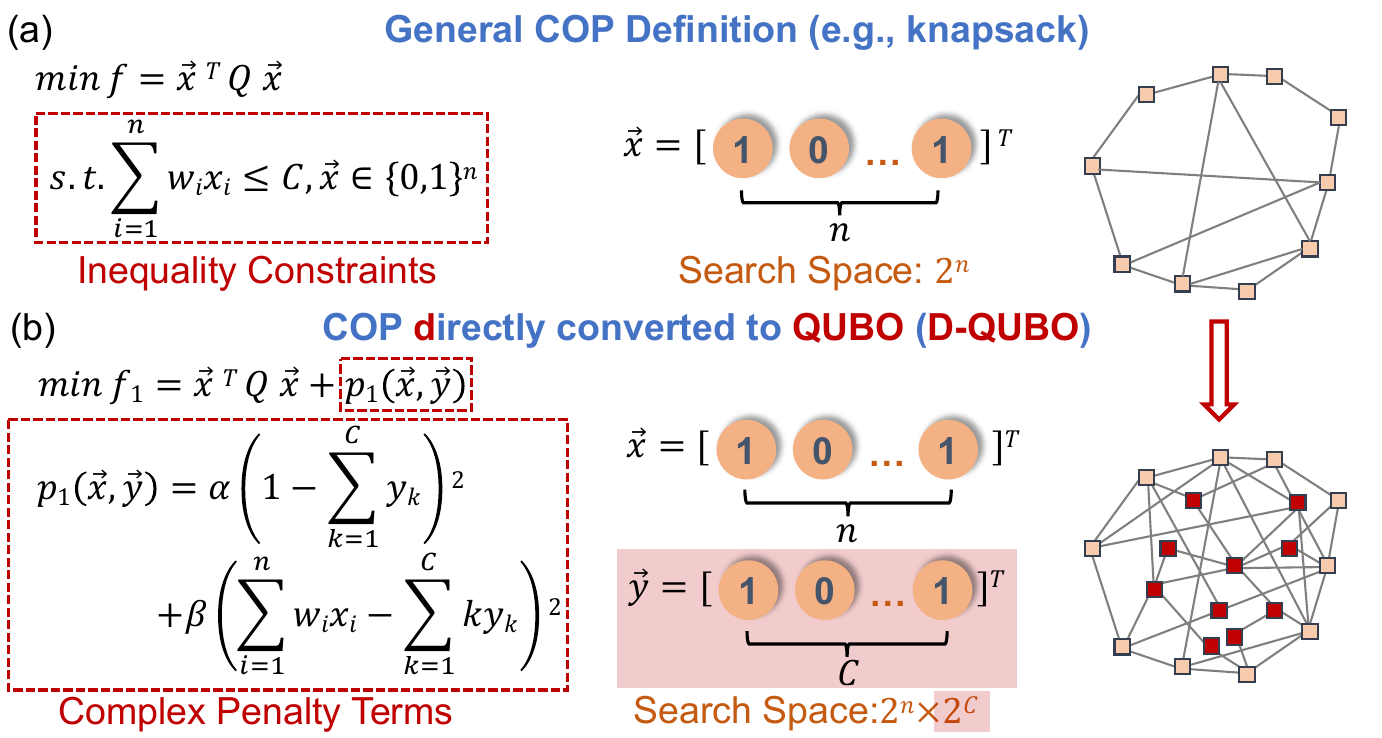}
  \vspace{-3.5ex}
  \caption{ {\bf (a)} A general COP with an inequality constraint involves a search space of $2^{n}$ variable configurations; {\bf (b)} D-QUBO introduces extra scaled auxiliary variable vector $\Vec{y}$, expanding the search space to $2^{n+C}$.}
  \label{fig:old mapping}
  \vspace{-5.5ex}
\end{figure}

It's worth noting that existing Ising machines and QUBO solvers cannot directly address COPs. The transformation of COPs into Ising models or their equivalent QUBO forms \cite{glover2018tutorial} is a prerequisite.
However, existing solvers exhibit certain limitations when transforming general COPs with inequality constraints into QUBO forms, such as (quadratic) knapsack and bin packing \cite{quintero2021characterizing}.
As Fig. \ref{fig:old mapping} shows,  traditional QUBO transformation 
directly embeds the inequality constraints as penalty terms in the objective function, referred to as "D-QUBO", introducing extra auxiliary variables and associated calculations.
Such direct transformation significantly expands the 
search space, and increases both the number and size of parameters in the objective function, leading to much slower solution convergence and larger hardware resources.
Due to the limited capacity of existing QUBO solvers and the  complexity introduced by D-QUBO, many COPs with inequality constraints remain unsolvable \cite{pusey2020adiabatic}.

To address aforementioned challenges, we introduce HyCiM, a hybrid computing-in-memory (CiM) QUBO solver framework that efficiently solves general COPs while minimizing hardware overhead and optimizing solving efficiency. The key contributions of this work are summarized as follows:
\begin{itemize}
    \item {A novel transformation method is proposed to transform a general COP with inequality constraints into an inequality-QUBO form. This conversion avoids introducing auxiliary variables and significantly reduces the complexity of the search space and objective function (Sec. \ref{subsec:transformation}).} 
    \item {We proposed "inequality filter", a ferroelectric FET (FeFET)-based CiM 
    architecture
    that accelerates inequality evaluations and filters out infeasible input 
    configurations by leveraging the multi-level characteristics 
    of FeFETs
    (Sec. \ref{sec:filter}).} 
    \item { We proposed a FeFET-based CiM annealer that approaches global solutions of COPs through iterative QUBO vector-matrix-vector (VMV) multiplication computations within a simulated annealing (SA) process (Sec. \ref{sec:crossbar}).} 
    \item {HyCiM achieved significant reduction in both search space and object function complexity, resulting in a higher COP-solving efficiency compared to other QUBO solvers (Sec. \ref{sec:eval}).} 
\end{itemize}

Our evaluation  clearly demonstrate that in solving 40 quadratic knapsack problems with 100 items,
HyCiM offers a substantial reduction in the search space (eliminating $2^{100} \text{ to } 2^{2536}$ infeasible input configurations) compared to the conventional D-QUBO approach. This reduction leads to significant hardware  overhead savings, ranging from 88.06\% to 99.96\%. 
Furthermore, HyCiM 
demonstrates high solving efficiency, achieving an impressive  success rate of 98.54\%, while D-QUBO implementation reaches only 10.75\%.

\vspace{-2ex}
\section{Background}
\label{sec:background}
\vspace{-1ex}
In this section, we first introduce the QUBO related concepts and works, then review FeFET based CiM basics.



\vspace{-1.5ex}
\subsection{QUBO basics and existing solvers}
\label{subsec:QUBO}
\vspace{-1ex}

Ising models offer a potent framework for modeling and solving COPs \cite{mohseni2022ising}, where problem variables are represented as spins, and  constraints are captured by spin couplings. 
The objective function of the problem translates  into the Hamiltonian energy function of the Ising model. Problem solving entails finding the spin configuration that minimizes the corresponding Ising Hamiltonian $H_{\mathrm{P}}$:
\vspace{-1ex}
\begin{equation}
\vspace{-1ex}
\label{equ:Ising model}
    \small \min H_{\mathrm{P}}=\sum_{i,j=1}^{N}J_{ij}\sigma_{i}\sigma_{j}+\sum_{i=1}^{N}h_{i}\sigma_{i}
\vspace{-0.5ex}
\end{equation}
where $N$ denotes the number of spins and $\sigma_i \in\pm1$ represents the state of spin $i$. $J_{ij}$ and $h_i$ stand for the coupling between spin $i$ and $j$ and the self-coupling of spin $i$, respectively.

The Ising Hamiltonian (Eq. \eqref{equ:Ising model}) is equivalent to  the QUBO form \cite{glover2018tutorial}  through a simple variable transformation $\sigma_i = 1 - 2x_i$, $x_i \in \{0,1\}$.
The QUBO form can be elegantly expressed as:
\vspace{-1ex}
\begin{equation}
\vspace{-0.5ex}
\label{equ:QUBO}
 \small \min y = \Vec{x}\ ^TQ\Vec{x}
\vspace{-0.5ex}
\end{equation}
where $\Vec{x} \in \{0,1\}^n$, and $Q$ is an $n \times n$ matrix.


While certain COPs, 
e.g., Max-Cut, graph coloring, 
can be seamlessly converted into QUBO, 
more general COPs with inequality constraints, e.g., (quadratic) knapsack, bin packing, etc \cite{quintero2021characterizing}, pose challenges for direct   QUBO transformation.

Fig. \ref{fig:old mapping}(a) depicts a general COP with an inequality constraint. 
The inherent search space of the problem is defined by the dimension of variable 
$\Vec{x}$, specifically $2^n$. 
Fig. \ref{fig:old mapping}(b) illustrates the D-QUBO transformation of the COP in (a) \cite{quintero2021characterizing, bontekoe2023translating}, 
which introduces an auxiliary variable $\Vec{y}$ and an extra penalty function $p_1(\Vec{x},\Vec{y})$ to represent the impact of inequality constraint. 
The first term in $p_1(\Vec{x},\Vec{y})$ tends to guarantee that $\Vec{y}$ has only one element equal to 1 and $\sum_{k=1}^{C}ky_k$ lies within the range of $\{1, 2, ..., C\}$, while the second term tends to ensure that 
$\sum_{i=1}^{n}w_ix_i = \sum_{k=1}^{C}ky_k$
, thus satisfying the inequality constraint.
The auxiliary variable $\Vec{y}$ and penalty function $p_1(\Vec{x},\Vec{y})$ 
significantly expand the search space from $2^n$ to $2^{n+C}$ and increase the number and size of parameters.
Addressing such D-QUBO form requires large-scale Ising machine and QUBO solver, resulting in higher hardware overhead and degraded solution convergence.


Numerous Ising machines and QUBO solvers have been developed to tackle D-QUBO forms.
Digital ASIC annealers, implementing diverse annealing algorithms to solve Ising models 
\cite{katsuki2022fast,takemoto20214}, 
may consume substantial energy and time  due to the exponential increase in explicit computations during each annealing iteration.
Dynamic system Ising machines 
leverages the inherent nature of physical systems to explore optimal solutions 
\cite{mallick2023cmos, afoakwa2021brim}, 
but they are highly sensitive to spin coupling implementations, that even slight deviations in coupling strength can disrupt convergence
\cite{ahmed2021probabilistic, mallick2023cmos}.
Moreover,  these Ising solvers are limited to COPs without self-interaction terms, such as  Max-Cut and Sherrington-Kirkpatrick models, as integrating self-interaction into these solvers is  non-trivial
\cite{moy20221, hamerly2019experimental}.
In this work,  we propose an alternative inequality-QUBO  implementation to  enhance the solving efficiency while avoiding the challenges associated with D-QUBO forms.

\begin{figure}[!t]
  \centering
  \includegraphics[width=1\columnwidth]{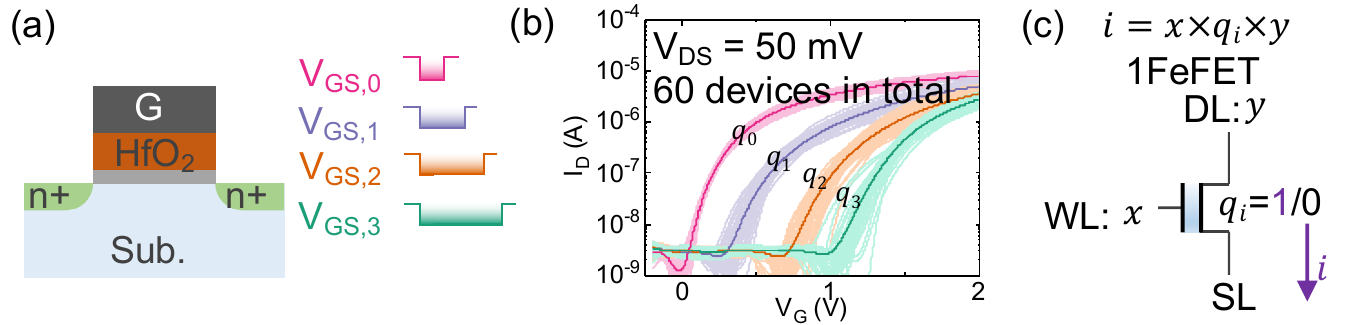}
  \vspace{-5.5ex}
  \caption{ 
  {\bf (a)} By applying different write pulses,
  {\bf (b)} multi-level $I_D$-$V_G$ characteristics of FeFETs can be programmed storing $q_0$ to $q_3$.
  {\bf (c)} FeFET storing a binary bit can naturally perform consecutive multiplications
  $i = x\times q_i \times y$ by simultaneously applying the inputs $x$ and $y$ to gate and drain, respectively.
  }
  \label{fig:FeFET device}
  \vspace{-5ex}
\end{figure}

\vspace{-1.5ex}
\subsection{CiM Preliminaries and FeFET Basics}
\label{subsec:CiM Preliminary}
\vspace{-1ex}

Prior  CiM implementations  employs non-volatile memories (NVMs) such as resistive random access memory (ReRAM) and magnetic tunneling junction (MTJ) to build crossbars   
that store  weight matrix, and perform parallel vector-matrix multiplication (VMM) operations for neural network accelerations 
\cite{zhao2024convfifo, mondal2018situ}.
However, these NVMs still face challenges for enhancing energy and area efficiency due to the current-driven mechanism, low ON/OFF ratio and large driving transistors.
FeFET, which integrates HfO$_2$ as the ferroelectric dielectric within the gate stack of MOSFET, has  emerged as a promising candidate for embedded memory and CiM.
Compared with  ReRAM \cite{salahuddin2018era} and MTJ \cite{zhuo2022design}, FeFET offers  better energy and area efficiency due to the  CMOS compatibility, voltage-driven mechanism, high ON/OFF ratio and three-terminal structure \cite{ni2019ferroelectric, yin2023ultracompact, hu2021memory, yin2022ferroelectric}.

When applied with different gate pulses as shown in Fig. \ref{fig:FeFET device}(a) at gate, 
a FeFET can store multi-level
$I_D-V_G$ curves which are experimentally measured from 60 devices as shown in Fig. \ref{fig:FeFET device}(b). Such multi-level characteristic can be utilized for our proposed FeFET-based CiM inequality filter.
When storing binary bit $q_i$, 
 a FeFET can 
inherently  perform multiplication $i = x \times q_i \times y$ through  drain-source current $i$ by applying  input signals $x$ and $y$ to  gate WL and  drain DL, respectively, as shown in Fig. \ref{fig:FeFET device}(c). 
Such single transistor multiplication is well-suited for  the proposed  FeFET-based CiM  crossbar supporting QUBO computation.

\begin{figure}[!t]
  \centering
  \includegraphics[width=0.95\columnwidth]{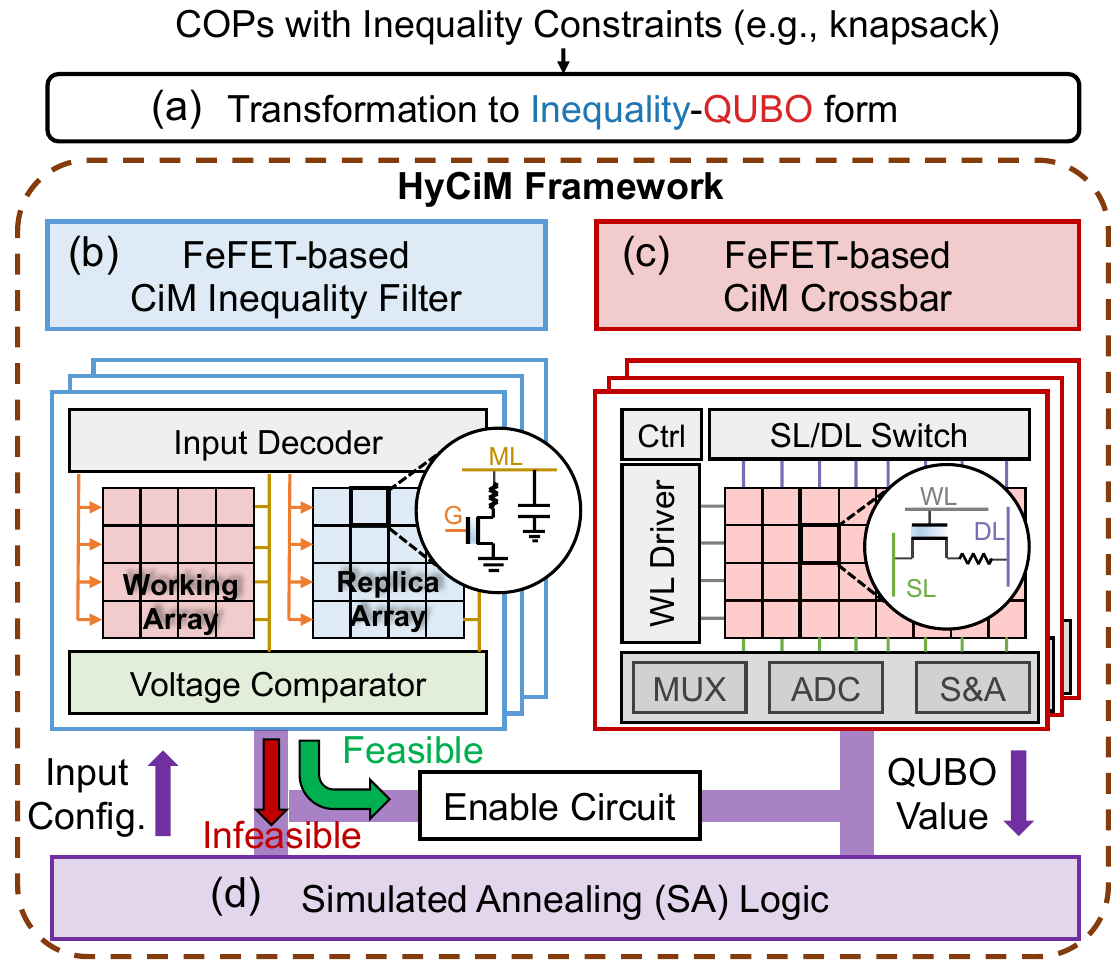}
  \vspace{-3ex}
  \caption{Overview of HyCiM. (a) Inequality-QUBO transformation method; (b) FeFET-based CiM inequality filter for  input configuration filtering; (c)(d) FeFET-based CiM annealer for approaching optimal solutions.}
  \label{fig:archi overview}
  \vspace{-5ex}
\end{figure}
\vspace{-2ex}

\section{HyCiM framework}
\label{sec:design}
\vspace{-1ex}

In this section, 
we  introduce our proposed HyCiM, including core components and operational flow. 

\vspace{-1.5ex}
\subsection{HyCiM Overview}
\label{subsec:archi overview}
\vspace{-1ex}

Fig. \ref{fig:archi overview} illustrates the overview of the proposed HyCiM framework. 
To tackle COPs with inequality constraints, e.g., knapsack, HyCiM first transforms these problems into an inequality-QUBO formulation (Sec. \ref{subsec:transformation}). 
The inequality-QUBO forms are then mapped onto 
a FeFET-based CiM inequality filter (Sec. \ref{sec:filter}) and a FeFET-based CiM Crossbar (Sec. \ref{sec:crossbar}), respectively. 
A SA logic is connected to these components to process the annealing (Sec. \ref{sec:crossbar}).

During each SA iteration, the SA logic generates an input variable configuration, which is then routed to the FeFET-based inequality filter. The inequality filter directly evaluates whether the configuration violates any constraints, and forwards feasible configuration to
the FeFET-based CiM crossbar for further QUBO computations. 
Infeasible configurations are returned to SA logic to generate  next input variable configuration.


\vspace{-1.5ex}
\subsection{Inequality-QUBO Transformation}
\label{subsec:transformation}
\vspace{-1ex}

COPs with inequality constraints represent a generalized category of COPs, as COPs without constraints or with equality constraints can be considered as special cases of COPS with inequality.
In this work, we focus on the general COPs with inequality constraints, and take quadratic knapsack problem (QKP) as the representative. QKP is a more general form of  knapsack problem, encompassing additional quadratic terms in the objective function \cite{quintero2021characterizing, bontekoe2023translating}: 
Given a set of items with weight and profit measurements, along with a knapsack limited in capacity, the objective is to select a subset of items that maximizes the total profit within the knapsack's capacity. 
Each item selected in the knapsack contributes an individual profit, and there are additional profits for pairs of selected items.
Mathematically, the problem can be stated as follows:
\vspace{-1.5ex}
\begin{equation}
\vspace{-1ex}
\label{equ:QKP}
 \small \max  \sum_{i,j=1}^{n}p_{ij}x_ix_j
\vspace{-1ex}
\end{equation}
\vspace{-1ex}
\begin{equation}
\vspace{-1ex}
\label{equ:QKP st}
 \small st. \sum_{i=1}^{n}w_ix_i\leq C, x_i \in \{0,1\}
\end{equation}
where $n$ is the number of items, $C \in Z^+$ is the capacity of knapsack, $w_i \in Z^+$ is the weight of item $i$, $p_{ij}(i \neq j)$ denotes the additional profit when $x_i$ and $x_j$ are selected, and $p_{ij}(i=j)$ denotes the individual profit of $x_i$. $p_{ij} = p_{ji}$. 
When item $i$ is selected, 
$x_i = 1$, otherwise $x_i = 0$.

In our proposed transformation, Eq. \eqref{equ:QKP} is  transformed to
\vspace{-1ex}
\begin{equation}
\vspace{-1ex}
\label{equ:QKP new}
 \small \min  f=\sum_{i,j=1}^{n}q_{ij}x_ix_j=\Vec{x}\ ^TQ\Vec{x}
\end{equation}
assuming $p_{ij} = -q_{ij}$.
Subsequently, rather than embedding the constraint  Eq. \eqref{equ:QKP st} directly into $f$ as D-QUBO (Fig. \ref{fig:old mapping}(b)), 
we reformulate the objective function as follows:
\vspace{-1ex}
\begin{equation}
\vspace{-1ex}
\label{equ:QKP final}
 \small \min  E=(\sum_{i=1}^{n}w_ix_i\leq C)\times \Vec{x}\ ^TQ\Vec{x}
\vspace{-0.5ex}
\end{equation}
The constraint  Eq. \eqref{equ:QKP st} is transformed to a logical function, and $E$ is non-positive. 
When Eq. \eqref{equ:QKP st} is satisfied, $E=\Vec{x}\ ^TQ\Vec{x}$, assuming a negative value to be minimized, otherwise, $E=0$.
This inequality-QUBO transformation separates the constraints from QUBO form, resulting in significant  search space reduction in QUBO variable configurations, and offering the potential for reduced hardware overhead and improved problem solving efficiency.

\begin{figure}[!t]
  \centering
  \includegraphics[width=1\columnwidth]{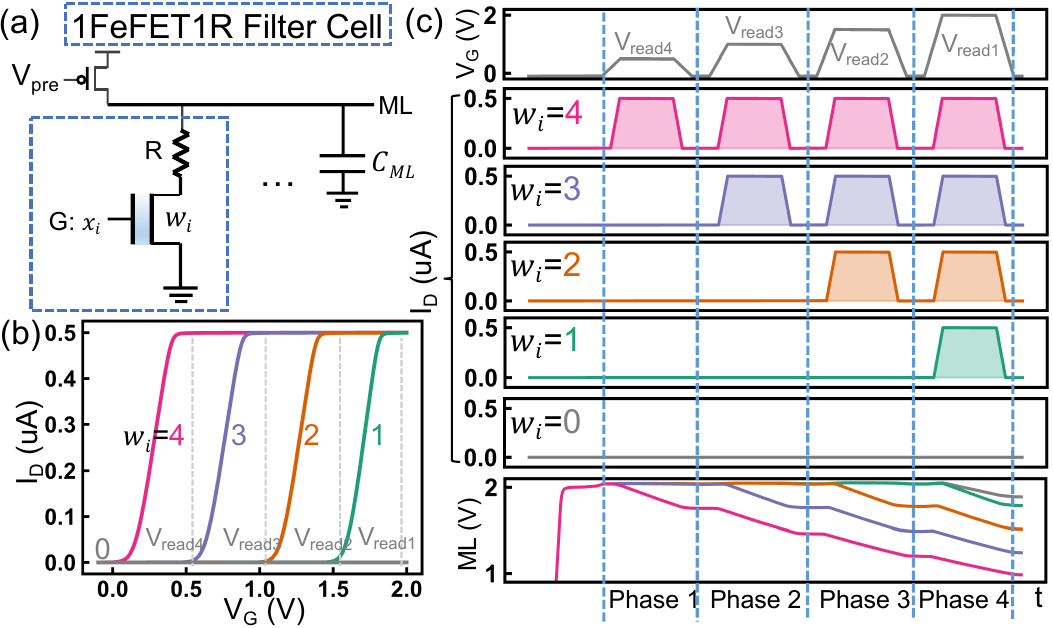}
  \vspace{-5.5ex}
  \caption{(a) Cell schematic of inequality filter array; (b) Transfer curves of cells storing five distinct weights and corresponding $V_{\text{read}}$'s; 
  (c) Transient waveforms of filter cells storing five  weights during the evaluation operation.
  }
  \label{fig:filter cell}
  \vspace{-5ex}
\end{figure}

\vspace{-1.5ex}
\subsection{FeFET-based Inequality Filter}
\label{sec:filter}
\vspace{-1ex}

Fig. \ref{fig:filter cell}(a) depicts the cell schematic of the FeFET-based CiM inequality filter array, comprising a single FeFET storing $w_i$ and a series resistor R.  
The input variable $x_i$ is applied at the gate. 
The cells within a word are connected to matchline (ML), with an additional capacitance $C_{\text{ML}}$ connected to ML. 
A precharging PMOS is connected to ML.
As depicted in Fig. \ref{fig:filter cell}(b), 
the cell's ON current is regulated by 1FeFET1R structure to reduce current variability \cite{soliman2020ultra, saito2021analog}  shown in Fig. \ref{fig:FeFET device}(b).
A cell can store five different weights $w_i=0 \text{ to } 4$, 
with four corresponding $V_{\text{read}}$ voltages. 
When a cell stores $w_i=k$, applying $V_{\text{readj}}$ 
($j \leq k$) 
turns the cell ON, conducting  $I_\text{D}$. Otherwise, the cell is OFF.

\begin{figure}[!t]
  \centering
  \includegraphics[width=1\columnwidth]{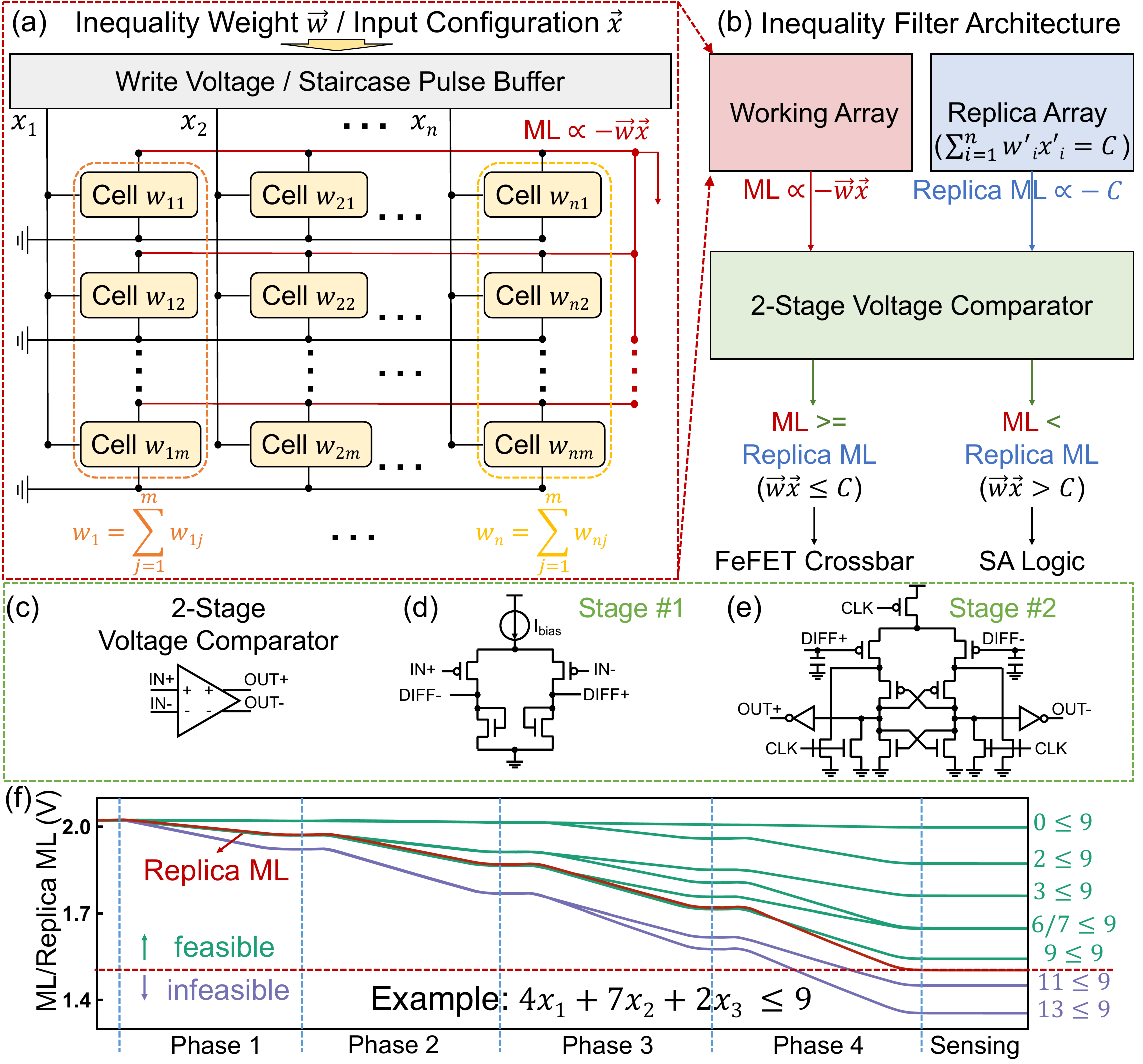}
  \vspace{-5.5ex}
  \caption{
  (a) Schematic of an $m \times n$ working array. The output $\text{ML} \propto - \Vec{w}\Vec{x}$; 
  (b) The inequality filter architecture, containing one working array, one replica array and a voltage comparator;
  (c) Symbol of  2-stage voltage comparator; (d)  First-stage differential pre-amplifier; (e) second-stage dynamic latched voltage comparator;
  (f) Transient  waveforms of an exemplary inequality evaluation. Infeasible configurations are filtered out.}
  \label{fig:filter array}
  \vspace{-5ex}
\end{figure}

Fig. \ref{fig:filter cell}(c) illustrates the cell operation of a filter iteration.
Firstly, ML is precharged to VDD (2V), followed by four phases.
During phases 1 to 4, if $x_i=0$, $V_\text{G}$ is  set to 0, and ML remains VDD.
If $x_i=1$,  a staircase pulse is applied at $V_\text{G}$, with  amplitude ranging from $V_{\text{read4}}$ to $V_{\text{read1}}$.
Only when $V_\text{G}$ is applied with $V_{\text{readj}}$ that $k \geq j$, can the cell turns ON, discharging ML.
As a result,  ML voltage $V_{\text{ML}}$ after a filter iteration can be expressed as follows:
\vspace{-1.5ex}
\begin{equation}
\vspace{-1ex}
\label{equ:1cell vdrop}
 \small V_{\text{ML}} = VDD -  k \times \frac{\int I_\text{D} t\ dt}{C_{\text{ML}}}  \times x_i
\end{equation}
By appropriately choosing  $C_{\text{ML}}$ and VDD, $\frac{\int I_\text{D} t\ dt}{C_{\text{ML}}}$ can be approximately  constant. 
Consequently, when $x_i=1$, $V_{\text{ML}}$ exhibits a linear relationship with $k$, which is also validated by simulation as shown in 
Fig. \ref{fig:filter cell}(c). 
Notably, Eq. \eqref{equ:1cell vdrop} is equivalent as below:
\vspace{-1ex}
\begin{equation}
\vspace{-0.5ex}
\label{equ:1cell algorithm}
 \small \text{ML} \propto - w_i x_i 
\vspace{-0.5ex}
\end{equation}

Building upon Eq. \eqref{equ:1cell algorithm}, we introduce an $m \times n$ working array for inequality evaluation, as depicted in Fig. \ref{fig:filter array}(a).
In the array, cells within a  column (e.g., column $i$) share an  input signal $x_i$, and all MLs are interconnected. 
The precharge circuit and $C_{\text{ML}}$ are not shown for simplicity.
The item weight vector $\Vec{w} = [w_1, w_2, ..., w_n]$ is stored horizontally across $n$ columns.
Suppose that each cell can store $0,1,...,k$ distinct weights. 
Accordingly, each item weight $w_i$ is decomposed into multiple $w_{ij}$ values, such that $w_i = \sum_{j=1}^{m}w_{ij}, w_{ij} \in \{ 0,1,...,k \}$, subsequently  stored in the cells within the $i^{th}$ column.
Eq. \eqref{equ:1cell algorithm} describes  ML of a single column, and  the summed ML  of the $m \times n$ working array can be expressed as:
\vspace{-1ex}
\begin{equation}
\vspace{-1ex}
\label{equ:working array algorithm}
 \small \text{ML} \propto - \sum_{i}^{n} w_i x_i 
\vspace{-0.5ex}
\end{equation}

Fig. \ref{fig:filter array}(b) shows the inequality filter, comprising a working array,
a replica array, and a 2-stage voltage comparator.
The replica array stores a precomputed weight vector $\Vec{w'} = [w'_1, w'_2, ..., w'_n]$, with a fixed input configuration $\Vec{x'}$ that satisfies $\sum_{i}^{n} w'_i x'_i = C$. The ML value of the replica  can be expressed as follows:
\vspace{-1.5ex}
\begin{equation}
\vspace{-1ex}
\label{equ:replica array algorithm}
 \small \text{Replica\ ML} \propto - \sum_{i}^{n} w'_i x'_i = - \text{C}
\vspace{-0.5ex}
\end{equation}

The two array ML outputs are fed to the 2-stage voltage comparator \cite{zhuo2022design} as shown in \ref{fig:filter array}(c-e) to compare $\sum_{i}^{n} w_i x_i$ and $C$.
If $\sum_{i}^{n} w_i x_i \leq C$, indicating a feasible input variable configuration, the FeFET-based crossbar  is activated to perform QUBO computation per Fig. \ref{fig:archi overview}. 
If 
$\sum_{i}^{n} w_i x_i > C$, indicating an infeasible configuration, a signal is transmitted to  SA logic, initiating  next SA iteration.

Fig. \ref{fig:filter array}(f) illustrates an  example where input variable configurations are evaluated and filtered for the inequality $4x_1+7x_2+2x_3 \leq 9$. 
The waveforms validate all possible input configurations ($2^3=8$ cases).
Following four phases, six MLs corresponding to feasible input configurations exhibit higher voltage than the Replica ML, 
while the other two MLs corresponding to infeasible input configurations drop below the Replica ML.

\begin{figure}[!t]
  \centering
  \includegraphics[width=0.95\columnwidth]{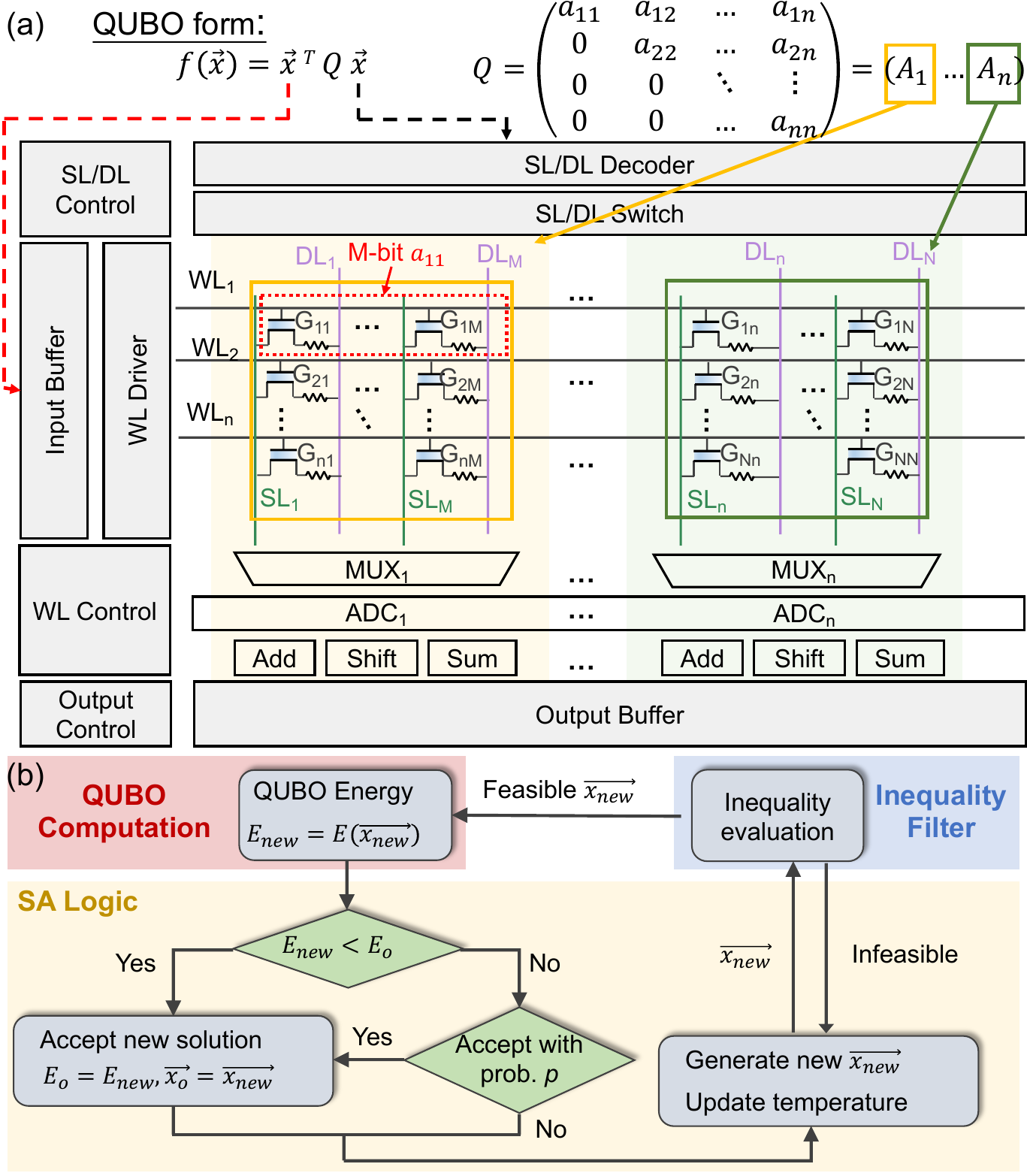}
  \vspace{-3ex}
  \caption{(a) The FeFET-based CiM crossbar array mapping the QUBO form;
(b) The simulated annealing flow, including Inequality filtering, QUBO computation and SA logic.}
  \label{fig:crossbar and SA}
  \vspace{-4.5ex}
\end{figure}

\vspace{-1.5ex}
\subsection{FeFET-based Crossbar and SA}
\label{sec:crossbar}
\vspace{-1ex}

Fig. \ref{fig:crossbar and SA}(a) shows our proposed FeFET-based CiM crossbar array that map  a general QUBO form $\Vec{x}^TQ\Vec{x}$ (Eq. \eqref{equ:QUBO}) by storing matrix Q and applying input vector $\Vec{x}$.
Each column of the matrix $Q$, denoted as $A_i$, is mapped onto an $n\times M$ subarray
, assuming $M$-bit matrix element quantization.
Each 1FeFET1R cell stores 1 bit of the element.
The feasible inputs  $\Vec{x}^T$ and $\Vec{x}$ are simultaneously directed to the input buffer and SL/DL decoder, respectively.
During each  QUBO computation, the drains and gates of FeFET devices are applied with corresponding binary bits of input vector $\Vec{x}$, and the column currents  representing the partial vector-matrix-vector multiplications  are sensed and accumulated together to get  the QUBO form result.

Fig. \ref{fig:crossbar and SA}(b) illustrates the SA flow, where the SA logic interacts with the crossbar and inequality filter.
During each SA iteration, the SA logic generates a new input configuration $\overrightarrow{x_{\text{new}}}$, which is then forwarded to the inequality filter for inequality evaluation.
If $\overrightarrow{x_{\text{new}}}$ is determined to be infeasible, a signal is sent back to the SA logic to initiate  next SA iteration.
If $\overrightarrow{x_{\text{new}}}$ is determined to be feasible, 
the inequality filter forwards $\overrightarrow{x_{\text{new}}}$ to the crossbar for QUBO computation. 
The resulting QUBO form output $E_{\text{new}}$ is then directed back to the SA logic, which  compares $E_{\text{new}}$
with  $E_{\text{o}}$ (the reserved output of the previous iteration)
and updates the optimal input configuration $x_{\text{o}}$ and corresponding QUBO form output $E_{\text{o}}$ per the comparison result and  probability associated with annealing temperature. 

\begin{figure}[!t]
  \centering
  \includegraphics[width=0.95\columnwidth]{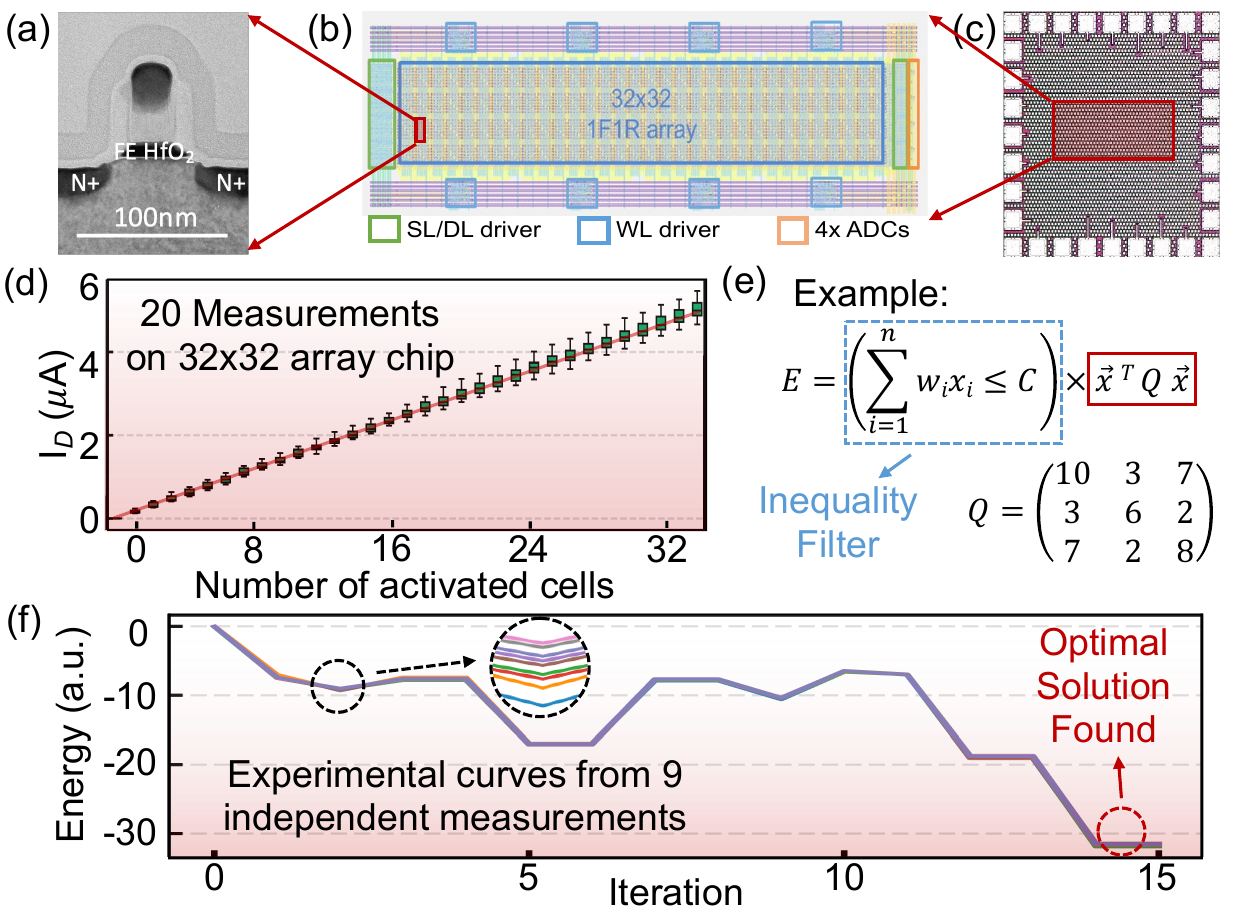}
  \vspace{-3ex}
  \caption{ 
  (a) TEM cross section of a FeFET; 
  (b) Layout of the $32\times 32$ FeFET array chip; 
  (c) Die photo;
  (d) Experimental measurements validating a good linearity of the crossbar;
  (e) The inequality-QUBO form of a QKP example;
  (f) Experimental energy evolution demonstrating successful problem-solving.
  }
  \label{fig:chip curve}
  \vspace{-4.5ex}
\end{figure}

Based on the structure in Fig. \ref{fig:crossbar and SA}(a), we fabricated a  $32\times 32$ FeFET-based  CiM array chip at 28nm high-k metal gate (HKMG) technology that implements the iterative QUBO computations.
Fig. \ref{fig:chip curve}(a-c) demonstrate the transmission electron microscopy (TEM) cross section of a FeFET device, chip layout
and the die photo, respectively.
Fig. \ref{fig:chip curve}(d) experimentally validates the current linearity with the  activated cell number within the $32\times 32$ array chip.
A QKP example  shown in Fig. \ref{fig:chip curve}(e) is solved by HyCiM, with annealing process validated on the chip.
Fig. \ref{fig:chip curve}(f) shows the  energy evolution with SA iterations under 9 independent experimental measurements.
For each measurement, the FeFET CiM array chip is erased and programmed again with the same QUBO matrix, and SA is executed. 
The curve results demonstrate that HyCiM successfully finds optimal solutions, affirming its capability and robustness 
in solving general COPs with inequality constraints.

\vspace{-2ex}
\section{Validation and Evaluation}
\label{sec:eval}
\vspace{-1ex}

In this section, we first validate the functionality of our proposed FeFET-based inequality filter and evaluate the hardware overhead reduction achieved by HyCiM in comparison to D-QUBO based annealer. 
We then evaluate the efficiency improvements in solving COPs with HyCiM compared to others.
All simulations were conducted using SPECTRE.
The Preisach FeFET model \cite{ni2018A} is adopted for FeFET, the 28nm TSMC model is used for MOSFETs with TT process corner at a temperature of 27$^{\circ}$C.
The wiring parasitics
are extracted from DESTINY \cite{poremba2015destiny}. 
We choose QKP as the representative COP and select 40 instances from \cite{QKPdataset}, each containing 100 items.

\vspace{-1.5ex}
\subsection{Validation of Inequality Filter}
\vspace{-1ex}
\label{sec:filter evaluation}

In our evaluation, both the working and replica arrays are of $16 \times 100$ size.
Each array column stores an item weight, ranging from 0 to 64.
As a result, the inequality filter effectively represents inequalities in the form of $\Vec{w}\Vec{x} \leq C$, $w_i \in \{0,1,...,64\}$.
For each  inequality of 40 QKP instances, we  generated 20 unique input configurations using  Monte Carlo sampling method, 
including 10 feasible ones and 10 infeasible ones, totaling 800 cases for evaluation. 

\begin{figure}[!t]
  \centering
  \includegraphics[width=0.95\columnwidth]{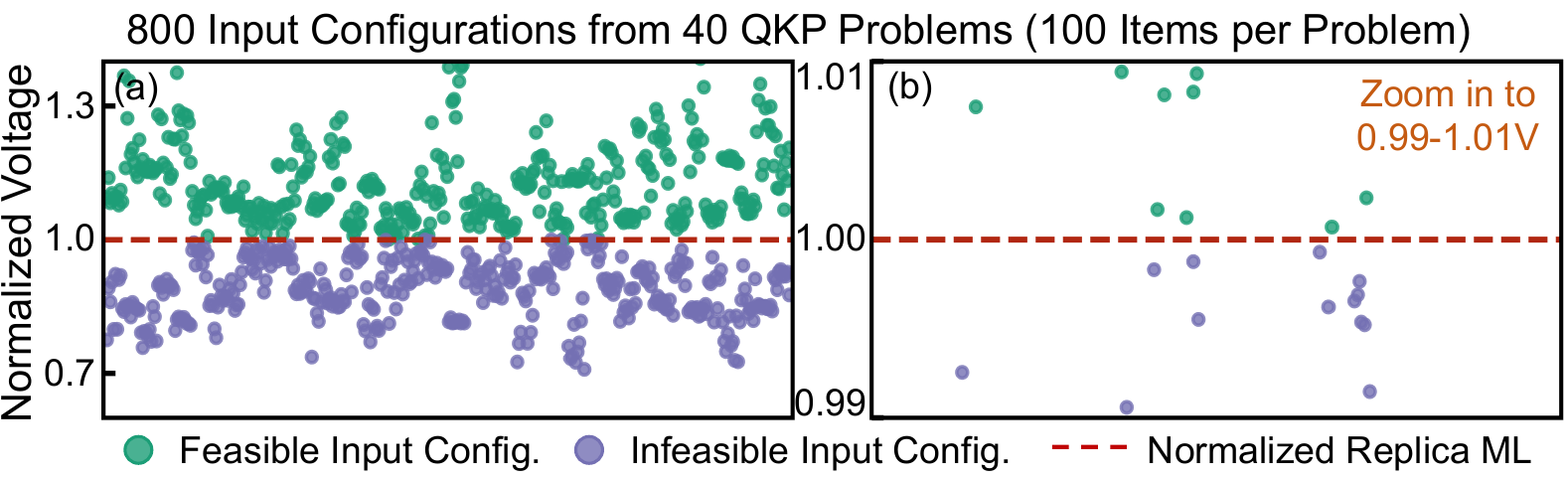}
  \vspace{-3ex}
  \caption{ 
  (a) Normalized ML outputs of the inequality filter classifying 800 input configurations for 40 QKP instances;
  (b) Closer view near the Normalized Replica ML.}
  \label{fig:dataset ml}
  \vspace{-3.5ex}
\end{figure}

Fig. \ref{fig:dataset ml}(a) displays the normalized ML output of working array over replica array for the 800 cases, where green circles represent feasible configurations, and purple circles infeasible ones. 
Fig. \ref{fig:dataset ml}(b) provides a closer view
near the normalized replica ML.
As can be seen, the inequality filter can clearly distinguish between feasible and infeasible configurations, whose  corresponding ML outputs
are distributed on the two sides of replica ML output, thus validating 
the effectiveness of the inequality filter.


\begin{figure}[!t]
  \centering
  \includegraphics[width=1\columnwidth]{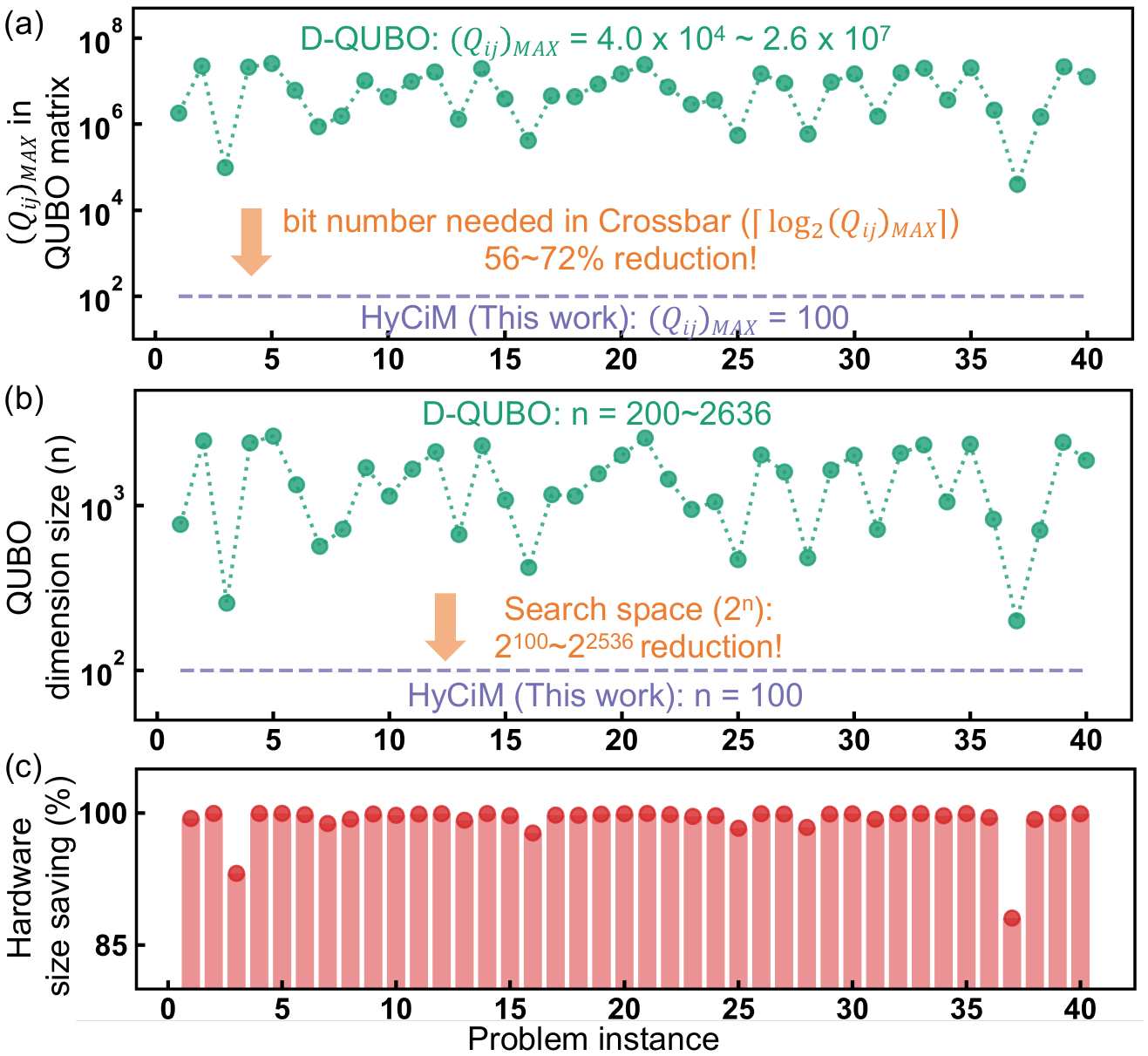}
  \vspace{-5ex}
  \caption{ 
  (a) The largest QUBO matrix element $(Q_{ij})_{MAX}$ in D-QUBO and HyCiM;
  (b) QUBO dimension size in D-QUBO and HyCiM; (c) Hardware size saving of HyCiM over D-QUBO.}
  \label{fig:hardware overhead}
  \vspace{-5ex}
\end{figure}

\vspace{-1.5ex}
\subsection{Hardware Overhead}
\vspace{-1ex}
\label{sec:hardware overhead}

We compare our proposed HyCiM  with D-QUBO approach in terms of hardware overhead. Both are implemented using the FeFET-based crossbar for fair comparison.
Coefficients $\alpha$ and $\beta$ in Fig. \ref{fig:old mapping}(b) are set to 2.
Each cell in the crossbar stores 1-bit.

When mapping the QUBO matrix $Q$ to crossbar, the quantization precision is determined by the largest matrix element, denoted as $(Q_{ij})_{\text{MAX}}$. 
Therefore, the crossbar needs $\lceil \log_2{(Q_{ij})_{\text{MAX}}} \rceil$ bits to  represent each element. 
In D-QUBO, the large-scaled auxiliary variable vector $\Vec{y}$ introduces a  large   $(Q_{ij})_{\text{MAX}}$ value. As shown in Fig. \ref{fig:hardware overhead}(a), the largest matrix element in D-QUBO approach can reach up to 4-7 orders of magnitude, 
which corresponds to 16-25-bit quantization in the crossbar.
On the contrary, HyCiM addresses the  constraints using inequality filter, thus $(Q_{ij})_{\text{MAX}}=100$, corresponding to 7-bit.
Therefore, the required quantization bits  are reduced by 56-72\% in HyCiM compared to D-QUBO.

In QUBO, 
the size of search space is  $2^n$, where $n$ refers to the size of QUBO matrix dimension, i.e., the size of binary input variable configuration. 
Fig. \ref{fig:hardware overhead}(b) illustrates the sizes of QUBO matrix dimension $n$ for both D-QUBO and HyCiM when processing 40 QKP problem instances. 
In D-QUBO shown in Fig. \ref{fig:old mapping}(b), $n$ equals to the total size of $\Vec{x}$ and $\Vec{y}$,
ranging from 200 to 2636, corresponding to a huge search space ($2^{200} \text{ to } 2^{2636}$).
On the contrary, in HyCiM, $n$ is 
100  since all instances contain 100 items, corresponding to a search space of $2^{100}$.
When compared to D-QUBO, HyCiM reduces the search space by $2^{100}\text{ to }2^{2536}$, effectively preventing unnecessary QUBO computations
and speeding up search space exploration.

When compared to D-QUBO, HyCiM not only reduces the required quantization precision  to store the QUBO matrix, but also significantly reduces the size of QUBO matrix dimensions, thus saving huge hardware overhead.
Fig. \ref{fig:hardware overhead}(c) illustrates the overall hardware size savings achieved by HyCiM (inequality filter and crossbar )over D-QUBO (crossbar only).
The results show that HyCiM  achieves hardware overhead reductions ranging from 88.06\% to 99.96\%, indicating
improved energy efficiency and performance, making HyCiM an appealing solution for COP solving COPs. 

\vspace{-1.5ex}
\subsection{Problem Solving Efficiency}
\vspace{-1ex}
\label{sec:problem solving}

\begin{figure}[!t]
  \centering
  \includegraphics[width=0.95\columnwidth]{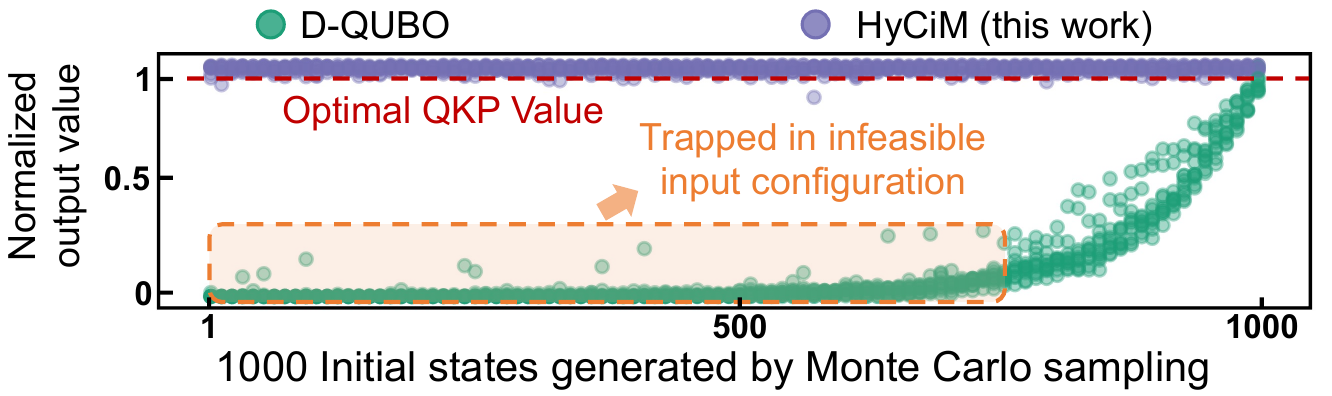}
  \vspace{-3ex}
  \caption{ 
    Normalized QKP values solved by D-QUBO based implementation and HyCiM.
    }
  \label{fig:SA result}
  \vspace{-4.5ex}
\end{figure}

We evaluate the QKP solving efficiency of 
D-QUBO based implementation and  HyCiM using FeFET based CiM.
We generate 1000 initial input
configurations for each  QKP instance using Monte Carlo sampling, and conduct 100 SA runs per initial configuration on both implementations to record the QKP values they can obtain. Each SA run executes 1000 iterations.
Fig .\ref{fig:SA result} shows the normalized QKP values found by both frameworks. The optimal QKP value is set as 95\% of the true optimal value. 
It clearly shows that HyCiM consistently finds solutions around optimal QKP value,
achieving a remarkable 98.54\%  success rate in average.
Conversely, D-QUBO mostly gets trapped with infeasible input configurations during SA, 
resulting in poor QKP values and a 10.75\% average success rate.


The solver summary in Table. \ref{table:evaluation} demonstrates that  the proposed HyCiM can 
solve more complex and larger general COPs with high efficiency than prior works by leveraging the proposed inequality-QUBO transformation to reduce the search space.

\begin{table}\huge
\caption{Summary of QUBO Solvers}
\vspace{-2ex}
\label{table:evaluation}
\centering
\resizebox{\columnwidth}{!}{
\begin{tabular}{| c | c | c | c | c | c| c|}
\toprule
\hline
Reference & \cite{cai2020power} & \cite{shin2018hardware}& \cite{hong2021memory}& \cite{yin2024ferroelectric}& \cite{taoka2021simulated} & This work\\
\hline
\multirow{2}{*}{COP}
 &  \multirow{2}{*}{Max-Cut} & Spin & Traveling &  Graph &  \multirow{2}{*}{Knapsack} & Quadratic\\
& & Glass & Salesman & Coloring & & Knapsack\\
\hline
Constraint  & - & - & Equality & Equality & Inequality & Inequality\\
\hline
Search Space & 
  \multirow{2}{*}{No} &
  \multirow{2}{*}{No} &
  \multirow{2}{*}{No} &
  \multirow{2}{*}{No} &
  \multirow{2}{*}{No} &
  \multirow{2}{*}{Yes} \\
Reduction & & & & & & \\
\hline
COP to QUBO & 
  \multirow{2}{*}{D-QUBO} &
  \multirow{2}{*}{D-QUBO} &
  \multirow{2}{*}{D-QUBO} &
  \multirow{2}{*}{D-QUBO} &
  \multirow{2}{*}{D-QUBO} &
  Inequality \\
Transformation & & & & & & -QUBO \\
\hline
Hardware Imp.$^\star$ & \multirow{2}{*}{Memristor} & \multirow{2}{*}{RRAM} & \multirow{2}{*}{RRAM} & \multirow{2}{*}{FeFET} & \multirow{2}{*}{RRAM} & \multirow{2}{*}{FeFET} \\
for Crossbar & & & & & & \\

\hline
Problem Size& 60 node  & 15 node  & 100 node & 21 node  & 10 node  & 100 node\\
\hline
Average & \multirow{2}{*}{65$^\dagger$}  &  \multirow{2}{*}{-} & \multirow{2}{*}{31$^\dagger$} & \multirow{2}{*}{-} & \multirow{2}{*}{92.4$^\dagger$} & \multirow{2}{*}{98.54}\\
Success Rate (\%) &   &   &  &  &  & \\
\hline
\bottomrule
\end{tabular}
\vspace{-1.5ex}
}

 \begin{flushleft}
 \scriptsize
$^\star$: Imp. denotes for implementation.
$^\dagger$: Extracted from literature.

\end{flushleft}
\vspace{-3ex}
\end{table}

\vspace{-2ex}
\vspace{-0ex}
\section{Conclusion}
\label{sec:conclusion}
\vspace{-1ex}


In this paper, we introduce HyCiM, a hybrid CiM QUBO solver framework designed to efficiently solve COPs with inequality constraints. 
We present a novel transformation method that converts COPs with inequality constraints into an inequality-QUBO form.
We propose a FeFET based inequality filter that evaluates inequalities and filters out infeasible inputs by leveraging the multi-level characteristic of FeFETs.
We fabricate a FeFET based CiM crossbar that accelerates the QUBO computations by exploiting the 
single transistor multiplication property of FeFET devices as well as SA process.
Evaluation results show that HyCiM offers substantial search space reduction, hardware overhead saving, and improved problem solving efficiency compared to prior QUBO solvers.
\vspace{-2ex}


\section*{Acknowledgment}

\vspace{-1ex}
This work was partially supported by NSFC (62104213, 92164203) and SGC Cooperation Project (Grant No. M-0612).
\vspace{-1.5ex}



%

\bibliographystyle{IEEEtran}
\bibliography{ref}


\end{document}